\documentclass[twocolumn,preprintnumbers,amsmath,amssymb]{revtex4}

\usepackage{graphicx}
\usepackage{dcolumn}
\usepackage{bm}
\usepackage{graphicx}

\begin{document}


\title{{\it Ab initio} simulation of market dynamics}

\author{Robert S. Farr}
 \affiliation{London Institute for Mathematical Sciences, Royal Institution, 21 Albemarle St, London UK, W1S 4BS}
 \email{robert.s.farr@gmail.com}

\begin{abstract}
We provide simple models for the utility function (or psychology) of an
actor trading a multitude of goods for money. In this framework, money has no
intrinsic consumption value, but is required as a medium of exchange. 
A collection of such actors are then
simulated interacting through market rules which create a double auction
for each of the goods. This framework captures the self-consistent, rational 
behavior of independent actors, including how they make compromises between 
purchases of different goods; so goes beyond price-demand curves, and 
also generates the small-scale fluctuations from individual trades.
We find that stable price formation requires a model that includes
time-preference for the actors. Fluctuations in prices show a distribution
with algebraic tails. Including inflation expectations leads to
complex, damped or un-damped price oscillations. 
We attempt to model the dynamics of input-output economic models, but
find it difficult to keep prices stable with the assumptions employed.
\end{abstract}

\maketitle

\section{Introduction}
The application of quantitative methods generally, and physically-inspired
models specifically, to economics has a long history \cite{Chen}.
Macro-economic models for the flow of goods and money have been built on
the basis of an analogy with hydraulic flows \cite{Phillips}. 
Kinetic theory of gasses has inspired models for unequal wealth distribution, 
for example in Angel's inequality process \cite{Angle,Chakraborti}.
In what can generically be described as `network theory' in statistical 
physics (and other fields), the development of preferential attachment models 
\cite{Mitzenmacher} to generate power law distributions of node degrees in 
graphs (or multiplicative processes to generate log-normal distributions)
has been used as a mechanistic model for both wealth distribution
and the size distribution of businesses.
The statistics of market fluctuations are also widely studied using
tools and models from statistical physics \cite{Plerou}.

In classical economics, input-output \cite{Leontief} models for a collection 
of industries can be used to predict required production to meet demand, 
for example in a planned economy. Classical theories of price formation 
are built on the concepts of supply-versus-price and demand-versus-price 
curves, where there is a 
unique equilibrium price that clears the market \cite{Hildenbrand}
(although, in close analogy with the Ising model \cite{Ising},
interactions between market participants can lead to distinct phases with 
different market prices \cite{Follmer}).

Despite the many physically-based models, the literature contains very few
models that could be described as {\it ab initio} in the sense that
a complete market is simulated from a fundamental model of the desires and 
constraints on simulated actors within the model. 
For example, input-output models may achieve an equilibrium when supply
meets demand, which makes supply and demand curves the 
foundation. Those demand curves assume,
usually (but not always \cite{Veblen}) correctly, that demand for a good
increases as its price falls. However, it is rarely taken into account that
an actual participant will have to make dynamic compromises when purchasing one
good among several in a functioning market. 

\begin{figure}
\begin{center}
\includegraphics[width=1.0 \columnwidth]{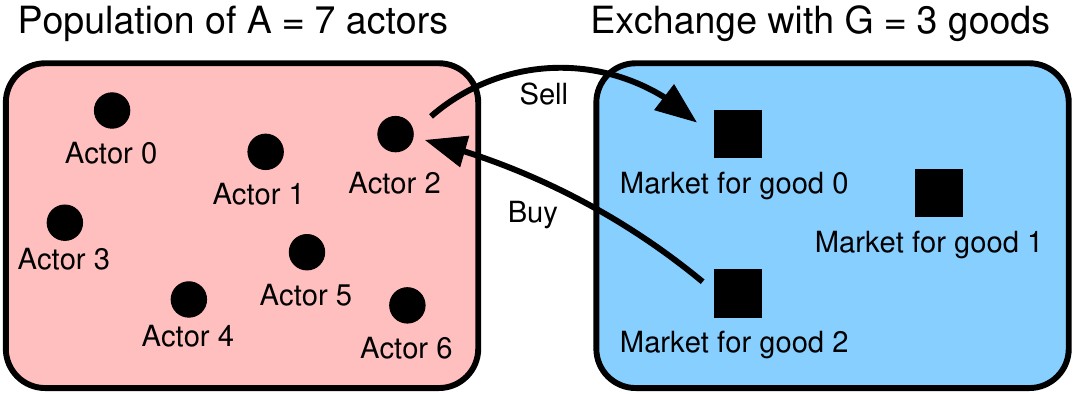}
\end{center}
\caption{A population of $A=7$ actors interacts with an exchange
of $G=3$ markets for individual goods. In this schematic, actor 2
is taking a turn in the markets, and is attempting to sell some
amount of good number $0$, and simultaneously buy some amount
of good number $2$.
\label{defs1}}
\end{figure}

Perhaps the clearest example
of an {\it ab initio} model in the sense meant here, is the minority game
\cite{Challet} (and various related problems \cite{Cavagna,Garrahan,Burridge}). 
In this model, it is the mental processes of the participants
that are simulated: each has a finite memory of past events, and chooses
her strategy based on how well different strategies would have worked
given the remembered past.

In the present work, we attempt to model the dynamics of a market (or
perhaps several markets) for multiple
goods from the rational behavior of a collection of market actors.
We do three things. First, we provide a set of market rules which must
be followed. These are designed to encourage trade and enforce the
orderly functioning of the market. Second, we provide a specific 
objective function for the actors. This can be extended in various ways
(as shown by example), but the key thing is that if the actors are attempting
to maximize a well-defined objective, this allows their behavior to be
self-consistently modeled when they are presented with a diversity of 
circumstances. Because of the complexity of the interaction between
market rules and the participants' conflicting desires, we also need
to provide (thirdly) some heuristics for how they attempt to reconcile
their behavior with those rules.

In each of these three parts, there is ambiguity, and so we will need to
make specific choices which could easily limit the generality of the work.
One may therefore ask why such a specific model is interesting, and in
particular why it might be of interest to physicists? On the subject of general
interest, there are many practically important concepts in economics, 
including (to take a few
random examples) velocity of money, inflation expectations and hyperinflation,
which might be clarified if they could be reproduced in a model where
every aspect of the processes is visible and could be analysed.
As far as physicists are concerned, one of the great insights from the
second half of the twentieth century was that the behavior of complex
systems can display universality, in the sense that some properties are
independent of detailed mechanisms, and instead are quantitatively
identical over large classes of models. Critical exponents, for example
of heat capacity, polarizability and fluctuations,
fall into universality classes in the neighborhood
of equilibrium phase transitions, and there are many more cases
for non-equilibrium phase transitions \cite{Lubeck}. 
This means that even specific and simplified
models can make quantitative predictions about real systems, governed
by the same underlying physics. Percolation \cite{Hammersley} is perhaps the 
most widespread example of this, but magnetic phase transitions and other 
changes of state have been productively analysed in the light of simple 
statistical models.

There is no guarantee that universality classes will emerge in economic 
behavior; but the existence of a complete and self-consistent model,
together with data from real life, may provide a basis for a search.

\section{The initial model}
\subsection{Definitions}
The model contains a number $G$ of goods, each referred to by an integer
$g\in\{0,1,\ldots, G-1\}$. For each of the goods there is a `market',
which is run as a double auction. There is a population of $A$ actors
(market participants), each referred to by an 
integer $a\in\{ 0, 1, \ldots,A-1\}$.
The actors must use `money', to buy and sell on the markets. Money is not
consumed, so it has no intrinsic value in terms of consumption; but because 
of the rule that it must
be used to buy and sell, it acquires some value as a medium of exchange.
Money is considered to be infinitely divisible; and again because of
the requirement to use it, it also functions as a unit of accounting
in the heuristics to be described below.

The collection of markets for different goods is called an `exchange'
(see Figure \ref{defs1} for a schematic). 
If we were to model long-distance
trades, actors would have access to more than one exchange, and furthermore
there may be more than one population taking part in the economy.
For the purpose of the models described here, we assume there is one population
and one exchange.

The time during which actors interact with an exchange is called a `trading 
period'. Actors take turns to interact with the exchange (under the market
rules to be described below, they are chosen randomly for a `turn').
We measure time during a trading period through `rotations'. A rotation
consists of $A$ actors having a turn in the exchange. The number of distinct 
actors taking a turn during a rotation will almost certainly be fewer
than $A$, due to the random nature of the choice [specifically, for
large $A$, we expect
approximately $(1-e^{-r})A$ distinct actors to interact with the markets
during the course of $r$ rotations, where $e$ is the base of the natural
logarithms].

\subsection{Market rules}
A market for each of the goods is run as a double auction. That means 
that a book of
buy orders and book of sell orders is maintained. The buy orders book is
a list of instructions from different actors to buy a specified amount of
that good, at a specified price. The money required to place a buy
order for such an actor $a_1$ is kept on escrow, and is not available 
for $a_1$ to make 
any other purchases, for as long as the order remains on the book. The highest
price buy order on the book is called the `bid price', $p_{{\rm bid},g}$.
Conversely, the sell orders book is a list of instructions from actors
to sell a specified quantity of this good for a specified price.
The volume of the good for sale by such an actor $a_2$ is kept on escrow,
and is accounted for slightly differently from the same amount of that good
which might remain in the possession of actor $a_2$. The lowest price
sell order on the book is called the `ask price', $p_{{\rm ask},g}$.

An actor $a_3$ entering the market for good $g$, and wishing 
to buy some amount of this good can either accept
(all or part of) any sell orders on the sell-orders book, or alternatively
place a new order on the buy-orders book.
Conversely, an actor wishing to sell some 
amount of good $g$ can either accept (all or part of) any buy order on the
buy-orders book; or alternatively, place a new order on the sell-orders book.
Only a finite number of orders are allowed on each book (we arbitrarily choose
$N_{\rm max}=20$ here), and if a new order is 
placed on one of the books which exceeds this
number, then an order from the opposite end of the book (the lowest price
buy order or the highest price sell order), is canceled, and the money
or goods (respectively) on escrow is returned to the actor who placed 
that order.

To encourage trade, there are simple rules on placing a new order
on one of the books. An actor placing a new order on the buy-orders book 
must do so at a less-favorable price than the current bid price.
Specifically, the new order must be at a price $(1+\epsilon)p_{{\rm bid},g}$,
where we choose $\epsilon = 0.01$. Conversely, anyone placing a new order on the
sell-orders book must do so at the less favorable
price $(1+\epsilon)^{-1}p_{{\rm ask},g}$. If the buy orders book becomes empty,
then we set the current bid price to the ask price for the same good, and
conversely. This leaves the only ambiguity that at the opening of the market,
the bid and ask prices are not currently defined, so they are both set
to an arbitrary value, which market forces will then correct. If the
buy orders book is empty, then an actor is allowed to place a new buy
order at a more favorable price, $(1+\epsilon_0)^{-1}p_{{\rm bid},g}$;
and an actor placing a new order on an empty sell-orders book may do
so at the more favorable price $(1+\epsilon_0)p_{{\rm ask},g}$,
where we arbitrarily choose $\epsilon_0 = 0.02$.

During a trading period, actors from the population are chosen randomly to
take a `turn' at the exchange. During this turn, the actor can attempt to
sell any amount and number of goods that he has in his possession
(i.e. not on escrow), subject to the condition that he does not have
an open buy or sell order on that good. He may also attempt to buy any amount
or number of goods for which he has the money in his possession (not on
escrow), again provided he does not have an open buy or sell order on the
books for that good.

\subsection{Objective function for an actor}
In a real exchange, actors might be wandering around eating 
(for example) hamburgers. 
However, in our simplified model, we strictly separate trade from consumption.
We imagine that an actor is undertaking trades on the exchange, with the
intention of accumulating goods which he will then conume after the
exchange closes.

For our first model of actor psychology, there is one period of market
participation, and then everything is consumed afterwards. The actor
therefore has no time preference in this model, and any money he is left with
after the market has no value as it cannot be consumed. For a single
population taking part in a single period of trade, this creates a frustration,
because everyone wants to spend all his money, but on average, everyone leaves
the trading period with the same amount of money he entered it with.
Section \ref{hyper} shows how this plays out in practice.
We will later revise this simple no-time-preference assumption, and
study the consequences.

The objective of an actor $a$ is to maximize his pleasure (or `utility')
from the consumption of 
all the goods he ends up with after the trading period. Our first model 
(`Model 1') for the psychology of the actor is that his utility function is,
\begin{equation}
\Omega_a^{[M1]} = 2\sum_g d_{a,g} \sqrt{x_{a,g}},
\label{Omega1}
\end{equation}
where $x_{a,g}$ is the quantity (or `volume') of good $g$ that actor $a$
ends up with (and consumes) following the trading period. The positive
value $d_{a,g}$ is the desirability of good $g$ to actor $a$, and initially
we take $\forall a,g: d_{a,g} = 1$.

If, at some point during the trading period, $a$ has a portfolio that consists
of the set of volumes $\{x_{a,g}\}$ of the different goods $\{ g \}$, together
with some money, $m_a$, then Eq.\ (\ref{Omega1}) allows $a$ to immediately
calculate the optimum portfolio he is aiming for. He does this by calculating
the amount of money $L_a$ he would achieve by liquidating (selling for 
money) his current portfolio at current market prices, and then calculating
the portfolio (the set of volumes $\{ y_{a,g} \}$ of goods) that he could 
buy with that money (at current market prices)
that would maximize $\Omega_a$. This is a straightforward calculation
optimizing over $\{ y_{a,g} \}$ with a Lagrange multiplier to keep
a constant liquidation value.
In this first model, that optimum
portfolio contains no money, but the optimum volumes of goods are
\begin{equation}
y_{a,g} = L_a \left.
\left(\frac{d_{a,g}^2}{p_g^2}\right)
\right/ \sum_{g'} \left( \frac{ d_{a,g'}^2 }{p_{g'}} \right),
\label{opt1}
\end{equation}
where $p_g$ is the current market price of good $g$. The consumption utility of
the optimum portfoilio is
\begin{equation}
\Omega_{a,{\rm opt}} = 2 L_a^{1/2}
\sqrt{ \sum_{g} \left( \frac{ d_{a,g}^2 }{p_g} \right) }.
\label{Omega1opt}
\end{equation}

We note two things. First, $\Omega_a$ is a convex function
of the volumes of goods, which captures the economic phenomenon of diminishing 
marginal utility. This property in some sense underpins all economic activity.
Second, because of the scale-free nature of the square root function,
the volumes of goods in the optimal portfolio for consumption, are proportional 
to $L_a$; in other words, their {\em relative} amounts do not depend 
on the liquidation value. This is a useful simplification for later, when
we consider phenomena such as time-preference and manufacture.

\subsection{Heuristics used by market actors}
The first thing an actor $a$, taking a turn in the exchange, needs to do, is
calculate what the liquidation value of his portfolio would be, at
current market prices. Market price is taken as the average of the bid and ask
price, which in practice are usually identical under the market rules 
(there is no `bid-ask spread'). The market value of a volume $x_g$ of
any good $g$ is the market price $p_g$ multiplied by $x_g$. 
The liquidation value of $a$'s portfolio is the sum of four
things: all the money he has in his possession (i.e. not on excrow);
the current market value of all the goods he has in possession (i.e. not
on escrow); the monetary value at the {\em order-book price} of any goods he 
has held on escrow on the buy-orders book (that is to say, the expected money
he will receive if the order is accepted); and {\em current} market value
({\em not} the money actually on escrow) of any goods he will receive from 
any orders on the buy-orders book.

Under the market rules, the actor $a$ is only allowed to trade in goods
for which he does not have a current order on the buy or sell book.
For goods which have a smaller volume in his optimum portfolio compared
to its volume in the current portfolio, actor $a$ sells as much as he
wishes to or can of each of these goods. The actor then prioritizes
goods he wants to buy (i.e. those whose volume in the optimum portfolio
is higher than in the current portfolio). For each such good, he calculates
the difference in volume of this good between the optimum and current portfolio,
and calculates the ratio of this to the volume in the optimum portfolio.
He then attempts to execute buy trades from highest to lowest values of
this ratio, until either he has tried all of them, or his money has run out.

This is the end of actor $a$'s `turn', after which a new actor is randomly
chosen for a turn in the exchange.

\begin{figure}
\begin{center}
\includegraphics[width=0.85 \columnwidth]{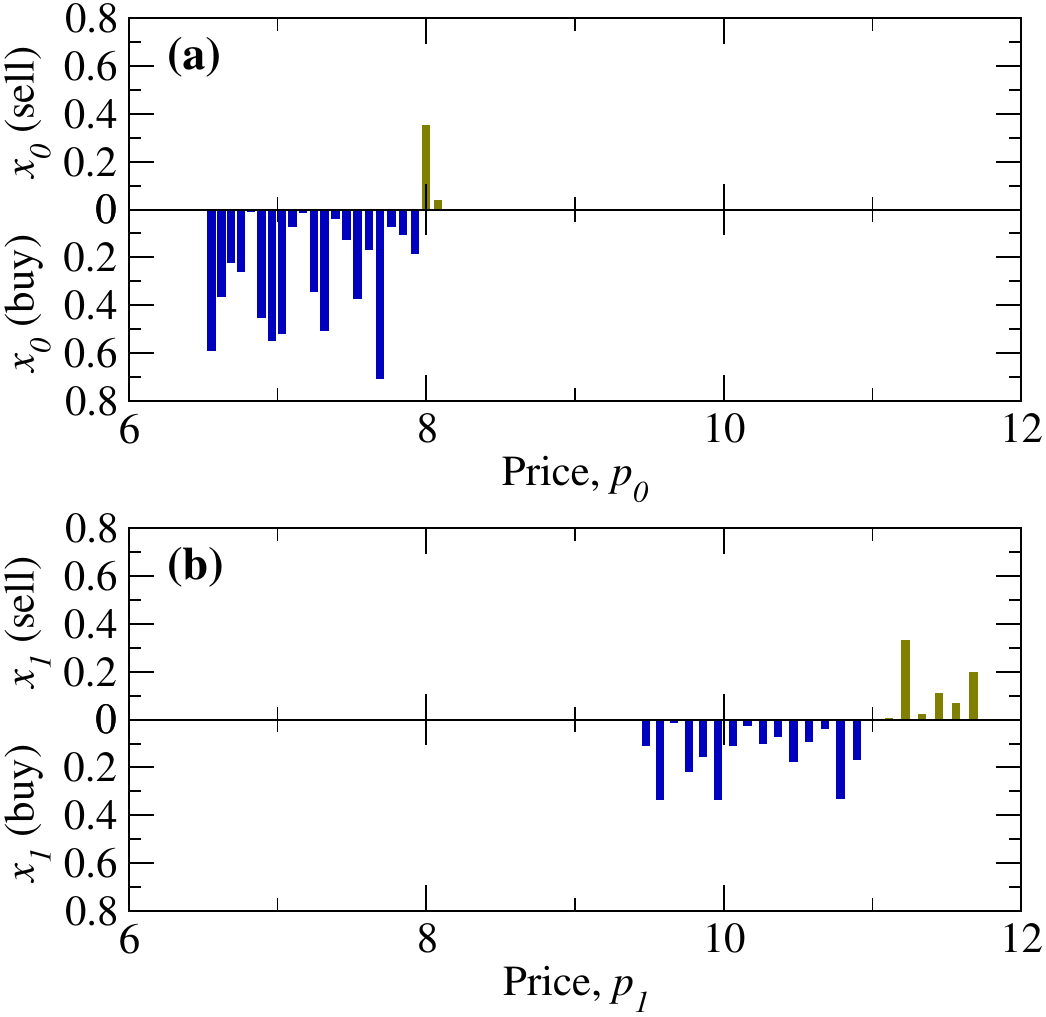}
\end{center}
\caption{Snapshot of the order books at turn 1000 for a population of 
$A=1000$ actors interacting with an exchange
of $G=2$ markets for individual goods. 
(a) Price and volume for open buy and sell orders for good $0$. 
(b) Open buy and sell orders for good $1$.
The initial (arbitrary) price of both goods is chosen to be 1.
\label{order_books}}
\end{figure}

\begin{figure}
\begin{center}
\includegraphics[width=0.85 \columnwidth]{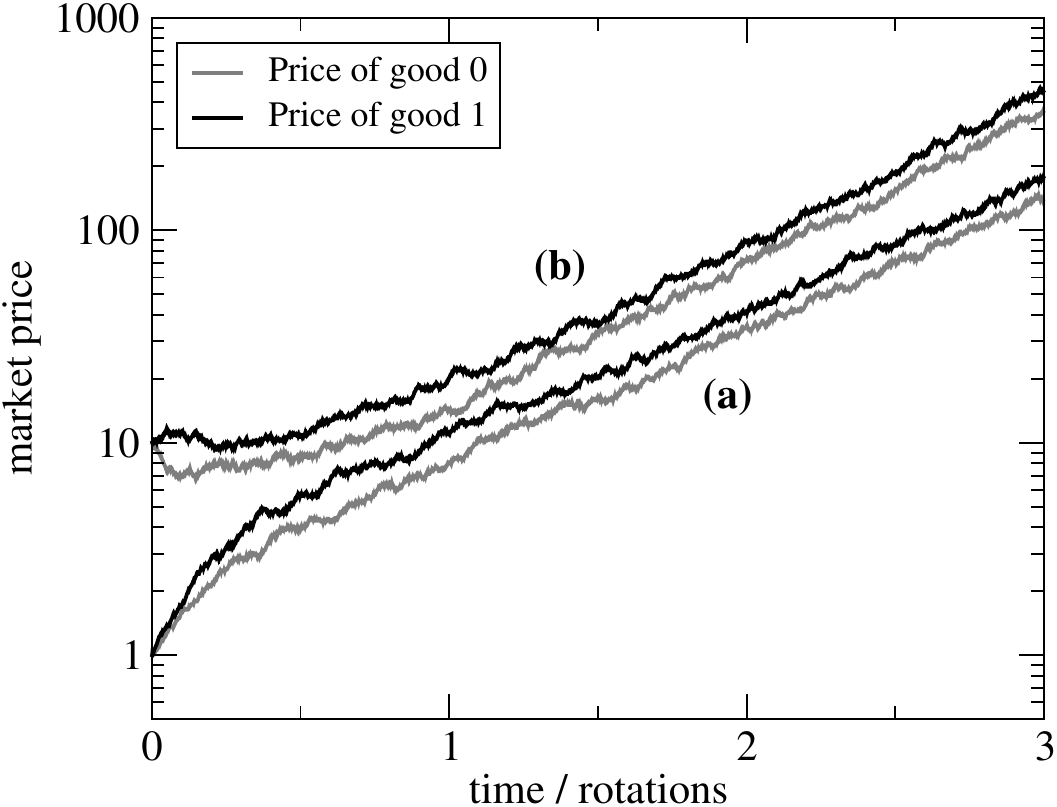}
\end{center}
\caption{Prices for goods when a population of $A=1000$ actors interacts 
with an exchange
of $G=2$ markets for individual goods. The prices of good $0$ is shown in gray,
and of good $1$ is shown in black. Two cases are shown: case (a) being
where the initial (arbitrary) price of both goods is chosen to be 1,
and case (b) where it is chosen to be 10.
\label{hyper_sim1}}
\end{figure}

\begin{figure}
\begin{center}
\includegraphics[width=0.85 \columnwidth]{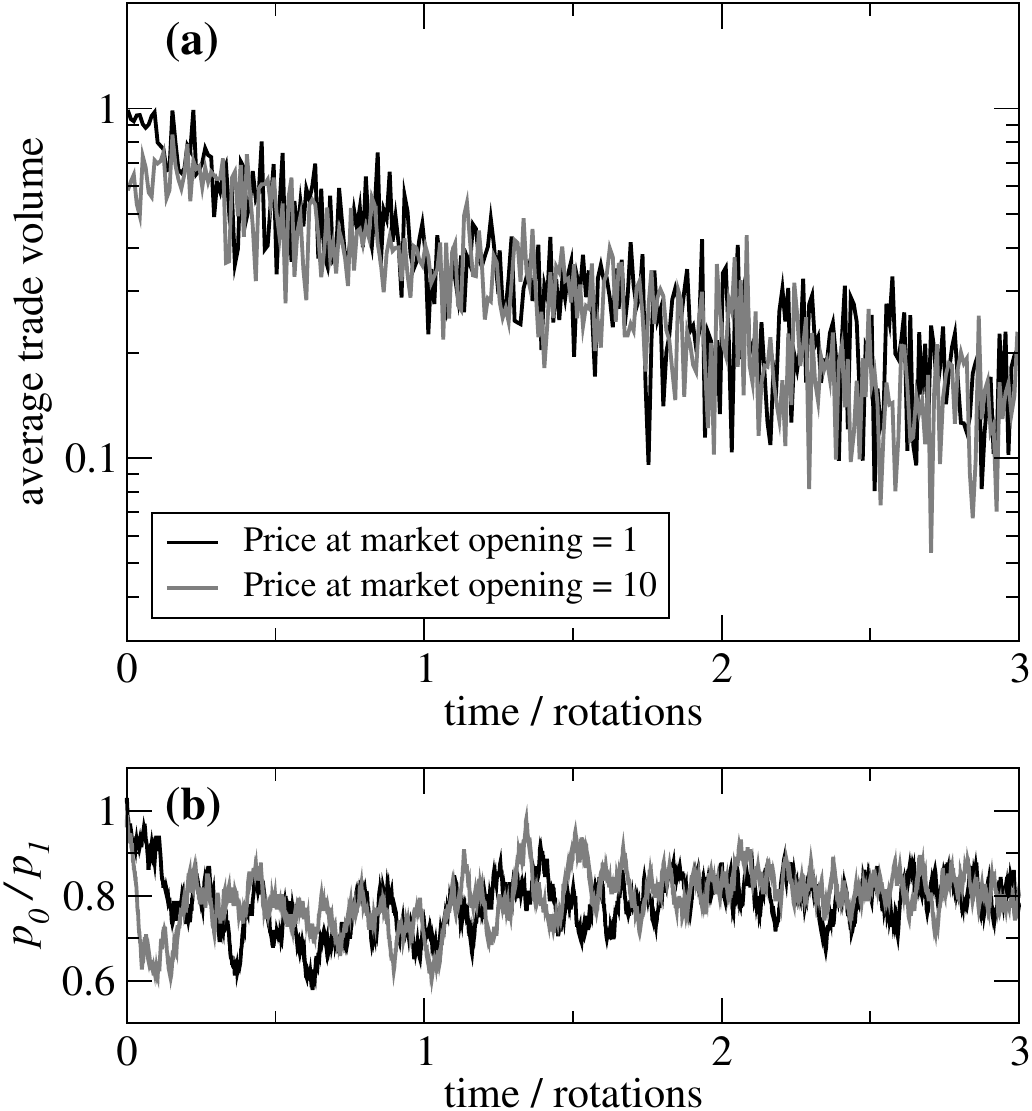}
\end{center}
\caption{(a) Total trade volume of goods (buy and sell), each point averaged 
over 10 turns. (b) The ratio of market prices for
two goods. In all cases, $A=1000$ actors interact
with an exchange of $G=2$ markets for goods. 
The initial prices of goods are arbitrary,and we show the cases where these
are chosen to be $1$ and $10$.
\label{hyper_sim2}}
\end{figure}

\section{Hyperinflation \& relative price\label{hyper}}
For the first model of actor psychology described here (`Model 1'), where
actors only consider what they will consume after a single trading period, 
money is only valued
implicitly as a means of transaction, rather than being a store of value.
Actors therefore wish to use money to buy valuable goods, but not to keep it.
In aggregate and on average however, they cannot escape owning the same amount
of money they had at the start of the trading period. How does this
frustrated system play out in practice? In figures \ref{hyper_sim1}
and \ref{hyper_sim2} we simulate a system with $A=1000$ actors, trading
$G=2$ goods. Each actor starts with $m=1$ unit of money. 600 of them
also start with one unit of good $0$, while the remaining 400
have one unit of good $1$. Figure \ref{hyper_sim1} shows the evolution
of the two prices, $p_0$ and $p_1$ through three rotations. The initial 
price, when all order books are empty, needs to be chosen arbitrarily,
and we show the two cases where the initial price for both goods is taken
as $1$ and $10$. Figure \ref{order_books} shows a snapshot of the order
books for the former case after 1000 turns (one rotation).

In figure \ref{hyper_sim1} we see the result is something that could
be described loosely as a hyperinflation, with prices rising exponentially
with time. The initial choice for prices does not appreciably affect the 
dynamics beyond the first rotation, other than introducing a factor into
both prices.

Inflation in the real world happens when liquidity grows faster than the 
production of goods and services. In our case, the quantity of money is fixed,
but the quantity of goods that are traded decays over time 
[see Figure \ref{hyper_sim2}(a)], as the actors gradually approach their 
optimum portfolios, and need to make smaller and smaller adjustments.
However, the underlying mechanism for inflation is closely analogous.

Despite the exponential rise in prices, the system does show evidence
of the formation of a stable {\em relative} price, as shown in
Figure \ref{hyper_sim2}(b). If the actors were allowed to trade by barter,
exchanging goods $0$ and $1$ directly, without using the medium of money,
there will be an exchange rate between the two goods which would match the
amount of goods up for sale from both types of actor (those that enter the
market with good $0$ and those that enter with good $1$). This could be 
described as the `market-clearing exchange rate'. To find this, consider
an actor $a_0$ who enters the market with unit volume of good $0$, and an actor
$a_1$ who enters the market with unit volume of good $1$. Let $\rho$ be the 
amount of good $1$ obtainable in the barter market for unit volume of
good $0$. Using Eq.\ (\ref{opt1}), we can map $a_0$ and $a_1$'s initial
portfolios $[x_{a,0},x_{a,1}]$ onto their optimal portfolios, 
$[y_{a,0},y_{a,1}]$:
\begin{eqnarray}
a_0 &:& (1,0) \rightarrow \left[ (1+\rho)^{-1},\rho^2(1+\rho)^{-1} \right],
\nonumber \\
a_0 &:& (0,1) \rightarrow \left[ \rho^{-1} (1+\rho)^{-1},\rho(1+\rho)^{-1} \right].
\nonumber
\end{eqnarray}
With $n_0 = 600$ actors like $a_0$ and $n_1 = 400$ actors like $a_1$, 
the amount of good $0$ (or good $1$) bought and sold matches, provided
\begin{equation}
\rho = \sqrt{n_1/n_0} \approx 0.816,
\label{eqm_price}
\end{equation}
which is consistent with the fluctuating ratio in Figure \ref{hyper_sim2}(b).

\begin{figure}
\begin{center}
\includegraphics[width=0.95 \columnwidth]{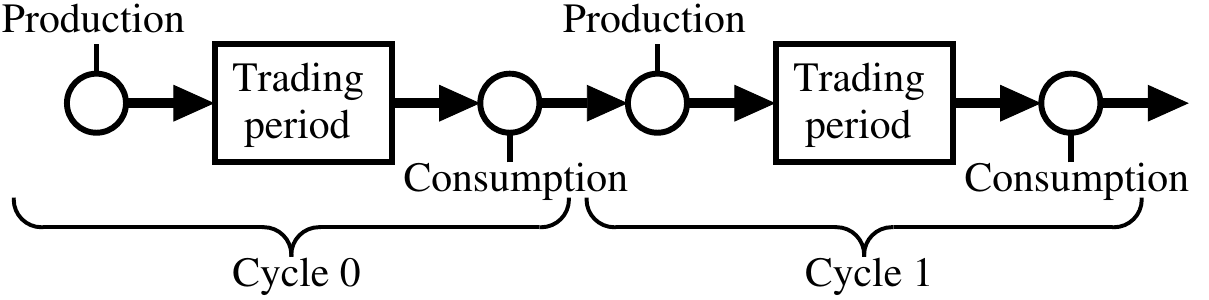}
\end{center}
\caption{A timeline of economic activity, which consists of a repeated
cycle of activity (two cycles being shown here). A cycle consists of
a period where actors produce new
goods for trade, followed by
a trading period, and finally  consumption, where the actors enjoy some or
all of their purchases (or unsold production). 
In following the dynamics, we will neglect the time taken for production
and consumption. In the second (and later) models of this paper, 
a trading period consists
of four rotations, and time will be expressed in `trading periods'.
\label{timeline}}
\end{figure}

\begin{table*}
\begin{tabular}{lllllll|cll|cll}
\hline
\hline
$A$ & $n_0$ & $n_1$ & $f_a$ & $\epsilon$ & $\epsilon_0$ & $N_{\rm max}$ & 
$\langle p_0\rangle$ & $\langle \ln p_0 \rangle$ & $\sigma_{\ln,0}$ & 
$\langle p_1\rangle$ & $\langle \ln p_1 \rangle$ & $\sigma_{\ln,1}$ 
\\
\hline
1000 & 500 & 500 & 1.5 & 0.01 & 0.02 & 20 &
$0.8024 \pm 0.0218$ & $-0.2205$ & $0.0272$ &
$0.8025 \pm 0.0218$ & $-0.2204$ & $0.0271$ 
\\
{\bf 200} & {\bf 160} & {\bf 40} & 1.5 & 0.01 & 0.02 & 20 &
$0.6635 \pm 0.0185$ & $-0.4105$ & $0.0279$ &
$1.3979 \pm 0.0581$ & $0.3341$ & $0.0411$ 
\\
{\bf 2000} & {\bf 1600} & {\bf 400} & 1.5 & 0.01 & 0.02 & 20 &
$0.6683 \pm 0.0168$ & $-0.4033$ & $0.0251$ &
$1.3418 \pm 0.0405$ & $0.2936$ & $0.0301$ 
\\
{\bf 20000} & {\bf 16000} & {\bf 4000} & 1.5 & 0.01 & 0.02 & 20 &
$0.6678 \pm 0.0171$ & $-0.4042$ & $0.0257$ &
$1.3368 \pm 0.0413$ & $0.2898$ & $0.0308$ 
\\
1000 & {\bf 600} & {\bf 400} & 1.5 & 0.01 & 0.02 & 20 &
$0.7359 \pm 0.0194$ & $-0.3070$ & $0.0264$ &
$0.9026 \pm 0.0252$ & $-0.1029$ & $0.0279$ 
\\
1000 & 500 & 500 & {\bf 2.0} & 0.01 & 0.02 & 20 &
$0.3339 \pm 0.0090$ & $-1.0973$ & $0.0269$ &
$0.3339 \pm 0.0090$ & $-1.0973$ & $0.0269$ 
\\
1000 & 500 & 500 & 1.5 & {\bf 0.02} & 0.02 & 20 &
$0.8046 \pm 0.0324$ & $-0.2183$ & $0.0402$ &
$0.8046 \pm 0.0324$ & $-0.2182$ & $0.0402$ 
\\
1000 & 500 & 500 & 1.5 & 0.01 & {\bf 0.05} & 20 &
$0.8023 \pm 0.0220$ & $-0.2206$ & $0.0274$ &
$0.8024 \pm 0.0220$ & $-0.2205$ & $0.0274$ 
\\
1000 & 500 & 500 & 1.5 & {\bf 0.02} & {\bf 0.05} & 20 &
$0.8045 \pm 0.0324$ & $-0.2183$ & $0.0402$ &
$0.8046 \pm 0.0325$ & $-0.2182$ & $0.0403$ 
\\
1000 & 500 & 500 & 1.5 & 0.01 & 0.02 & {\bf 50} &
$0.8026 \pm 0.0216$ & $-0.2202$ & $0.0269$ &
$0.8026 \pm 0.0216$ & $-0.2203$ & $0.0269$ 
\\
\hline
\hline
\end{tabular}
\caption{Statistics of price fluctuations between a time of 2.25 and 2.5 
trading periods,
from repeated simulations to accumulate $10^7$ price pairs. 
The numbers of actors who produce
(respectively) unit volume of goods $0$ and $1$ are $n_0$ and $n_1$. 
$N_{\rm max}$ is the maximum number of orders on an order book. 
$\sigma_{\ln,g}$ 
is the standard deviation of the natural logarithm of the price $p_g$.
All actors start with unit amount of money $m$. Bold face highlights which
parameters have changed compared to the first simulation in the table.
\label{fluctuation_params}}
\end{table*}

\begin{figure}
\begin{center}
\includegraphics[width=0.85 \columnwidth]{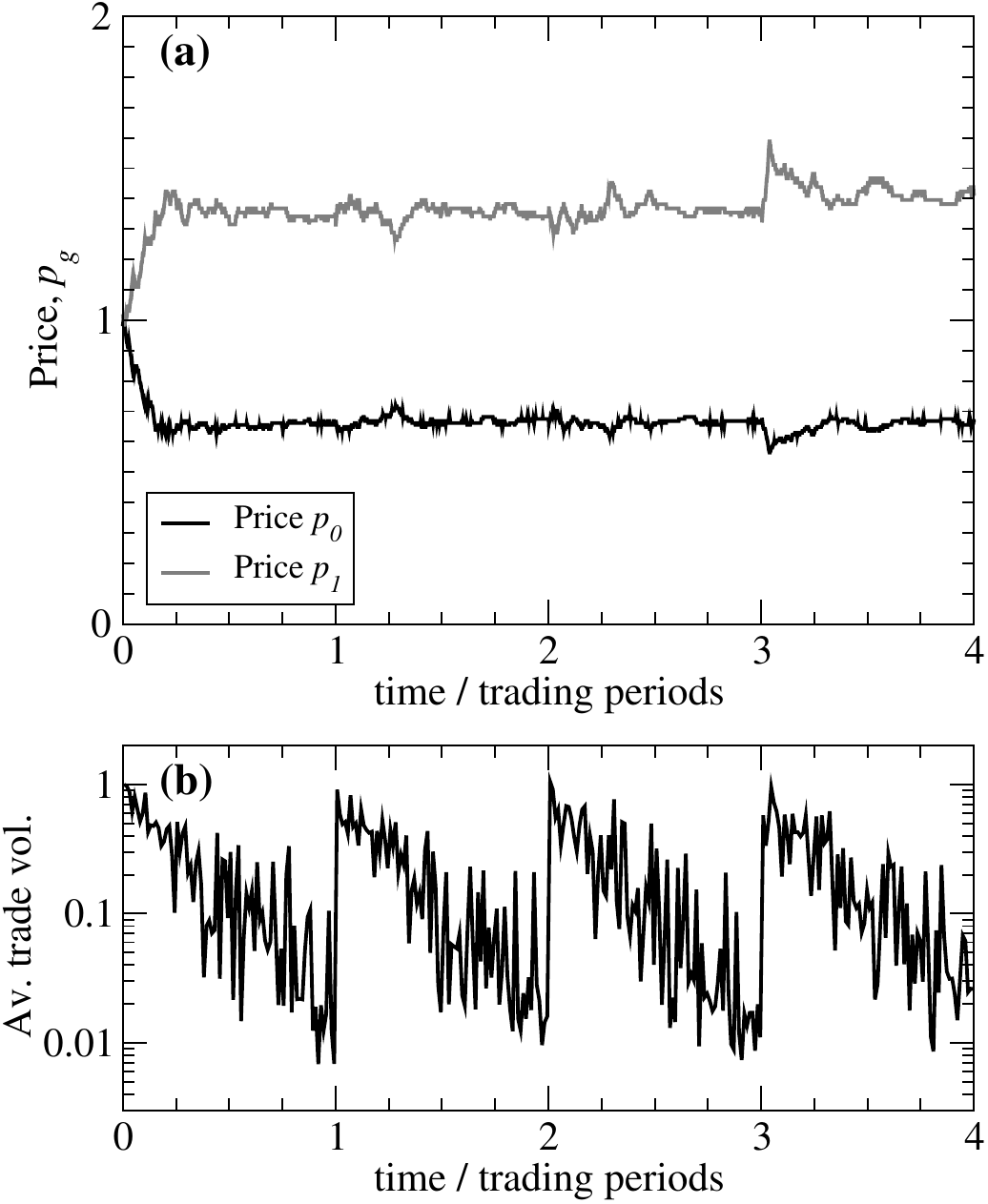}
\end{center}
\caption{(a) Prices $p_0$ and $p_1$ over the course of four trading periods, 
each such period
being four rotations. (b) The average of total trade volume
of goods (buy and sell) over 10 turns. 
In all cases, $A=200$ actors interact
with an exchange of $G=2$ markets for goods. 160 actors produce unit volume
og good $0$ each trading period, and 40 actors produce unit volume of good $1$.
All actors start with an amount $m=1$ of money, and all goods are equally
desirable, $d_{a,g}=1$.
The time-preference parameter is $f=3/2$.
\label{timepref_sim1}}
\end{figure}

\begin{figure}
\begin{center}
\includegraphics[width=0.85 \columnwidth]{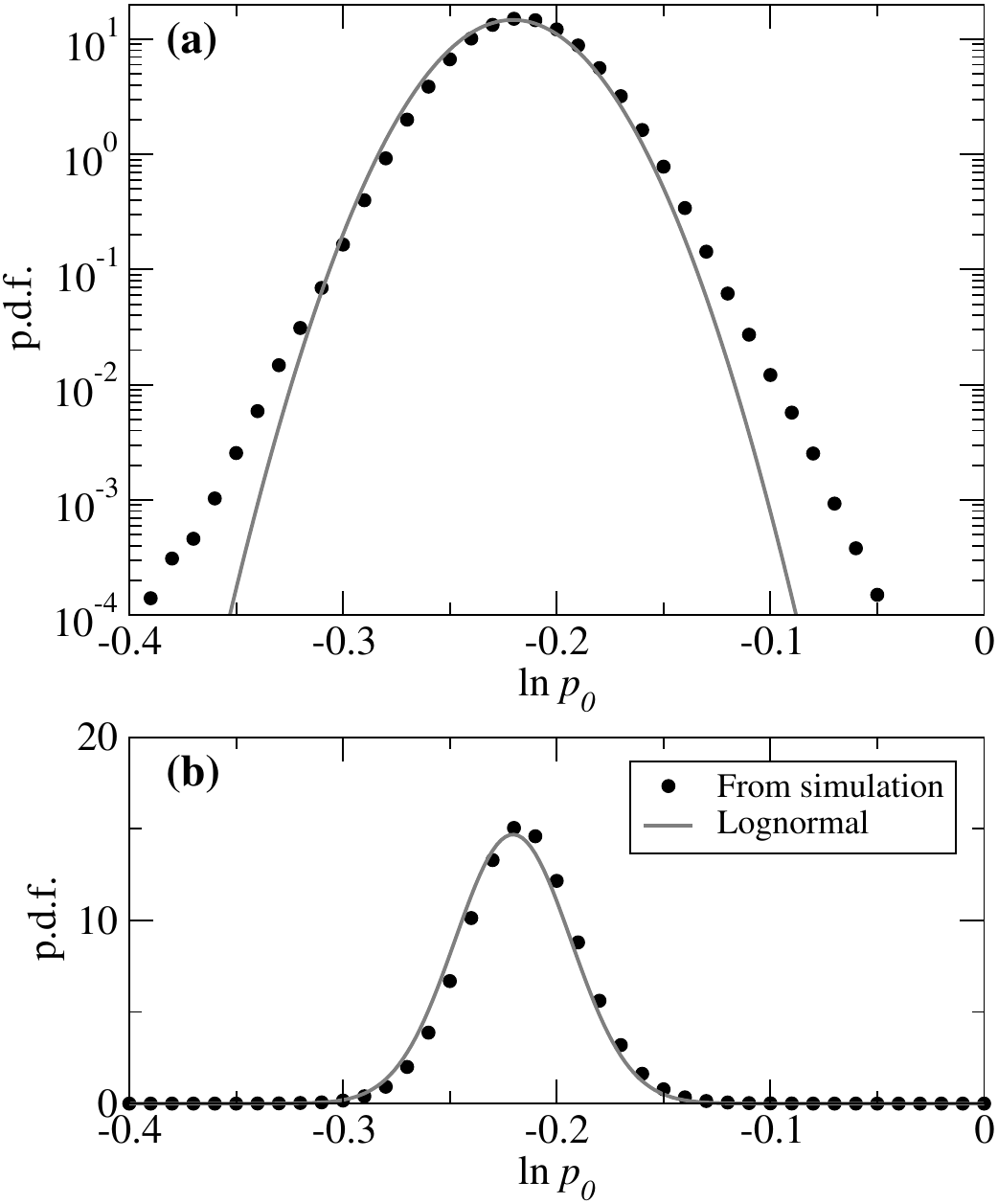}
\end{center}
\caption{Distribution of price fluctuations for the $10^7$ values of
$p_0$ in the first simulation of Table \ref{fluctuation_params}, together 
with a lognormal distribution with the same value of $\langle\ln p_0\rangle$
and $\sigma_{\ln,0}$. (a) Probability density function (p.d.f.)
plotted on a logarithmic scale. (b) P.d.f. plotted on a linear scale.
\label{log_pdf}}
\end{figure}

\section{Time preference \& stable prices}
So far, we have considered a single trading period, followed by consumption
of goods, which all the actors are planning for. We now define an extended
timeline, shown in figure \ref{timeline}, where there are many cycles of trading
periods, each preceded by a `production period' where actors produce new goods,
and followed by a `consumption period' where goods are consumed and their
utility realized. In our model, these three activities are absolutely separated.
Although, in reality, production and 
consumption take time, this is not of interest to us here, so we will
only follow time through the progress of the trading period. We therefore
treat a cycle metrically as a period of trading time, and in this
second model we take a trading period to be four rotations, to allow
most of the desired trading to take place for the vast majority of actors. 
A time of `1.5 trading periods' will therefore be the end of the 
second rotation in the 
second trading period. If the present time is $t$, then consumption
will formally take place at a time $\lceil t \rceil$, where we use the
`ceiling' notation $\lceil\cdot\rceil$ to denote the smallest integer greater
than $t$. At the end of a trading period, any outstanding orders on the
books are canceled and goods or money on escrow returned to their owners.

In order to give some intrinsic value to money, and so avoid the phenomenon
of hyperinflation during a trading period, for this second model, we posit that
all goods are perishable, and cannot be stored beyond the consumption period
following the current trading period. We also introduce a notion of time
preference into the psychology of the actors. We do this by modifying the
utility function so that an actor considers not only consumption which
immediately follows the current trading period, but also consumption
in the next consumption period. The result is `Model 2':
\begin{equation}
\Omega_a^{[M2]}(t) = 2\sum_g d_{a,g} \left[
\sqrt{x_{a,g}(\lceil t\rceil)} +
f_a \sqrt{x_{a,g}(\lceil t+1\rceil)}
\right] ,
\label{Omega2}
\end{equation}
where $t$ is the time in trading periods, and $f_a$ is a parameter 
controlling the actor's
future preference. Model 1 corresponds to $f_a = 0$, where actors take no 
interest in what happens after the consumption period following the current
trading period.

In this second model, the actor assumes prices are stable, so the current
market price can be used as an estimate for the price in the following
trading period as well. Let us assume that in the next production period,
the actor will produce goods of type $\gamma$ which have a market value 
of $P_a(\lceil t\rceil)$. This value is estimated from his production capacity
$c_a$ for good $\gamma$ (which is fixed, and usually taken here to be 1)
and the current market value of that good:
\begin{equation}
P_a = c_a \, p_\gamma.
\end{equation}
After calculating the current liquidation value
$L_a(\lceil t\rceil)$ of his portfolio, the actor will plan to consume
a fraction $(1-\phi)$ in the upcoming consumption period, leaving a value
$\phi\,L_a(\lceil t\rceil)+P(\lceil t\rceil)$ to be consumed in the
following consumption period at time $\lceil t+1 \rceil$.
The first term in this value, $\phi\,L_a(\lceil t\rceil)$, can 
only be carried through to the next trading period in the 
form of money, given the assumed perishability of all goods. The other
term, $P(\lceil t\rceil)$, is the value of goods he will produce
during the production period, and so the value is embodied in those goods.
Using Eq.\ (\ref{Omega1opt}), the utility once the actor has converted
liquidation value back into an optimal set of goods will be,
\begin{eqnarray}
\Omega_a^{[M2]} &=& 2
\left\{
\left[ (1-\phi) L_a\right]^{1/2}
+f_a \left[ \phi\, L_a +P_a \right]^{1/2}
\right\} \nonumber \\
&\times&
\sqrt{ \sum_{g} \left( \frac{ d_{a,g}^2 }{p_g} \right) } .
\label{Omega2phi}
\end{eqnarray}
Since, by assumption, $d_{a,g}$ and $p_g$ are fixed, the optimum value of
$\phi$ is
\begin{equation}
\phi_{\rm opt} = \left\{
\begin{array}{ll}
\frac{f^2 - (P_a / L_a)}{1+f^2} & P_a \le f^2 L_a \\
0 & P_a \ge f^2 L_a
\end{array}
\right. ,
\label{phiopt}
\end{equation}
which sets the amount $\phi_{\rm opt}L_a$ of money that the actor enters 
the next trading period with (as well as the goods he has produced in the 
production period).

We see that if $0\le f_a\le 1$, and the current liquidation value of the 
portfolio is equal to
the value which the actor will produce in the next production period,
then the actor will behave as in model 1, and consume everything he has
in the preent trading period immediately after. $L_a \approx P_a$ is likely
to be the case after stability is reached in the system, so the naively 
reasonable choice $0\le f_a\le 1$ does not introduce a significant 
time-preference into the system.

There are a number of ways to give the actor a significant time preference,
which will be necessary to ensure money has utility as a store of value.
For example, we may specify that the actor only produces during alternate
production periods, so he needs to make provision for (again alternate)
cycles when $P_a(\lceil t\rceil)=0$. The simpler method, which we adopt here,
is to give the actor a time preference parameter
$f_a>1$. This may not be unreasonable if planning for the later consumption
period is a proxy for considering the more distant future as well.
As a definite example in this second model, we will choose 
$\forall a:f_a = 3/2$.

Figure \ref{timepref_sim1} shows a simulation of $A=200$ actors over 4 
trading periods
of four rotations; 160 of whom produce unit amount of good $0$ before
every trading period, and
the remainder unit amount of good $1$. All $d_{a,g}$ are taken to be unity. 
From Eq.\ (\ref{eqm_price})
we therefore expect an equilibrium price ratio of $p_0/p_1 = 0.5$.
In fact, the average price ratio from time 1.0 to 4.0 trading periods 
is $0.480$ with standard
deviation $0.027$, which is consistent with the prediction.
Figure \ref{timepref_sim1}(b) shows the decay in trading volume during
each trading period
as actors get closer to reaching their optimum portfolios; however
at the end of each trading period, there is a consumption period and 
then a production period;
and actors enter the next trading period needing to trade again. During the
course of the trading period, the falling volume of goods does not lead
to inflation, because actors are also trying to acquire a non-zero amount
of money, so the amount of money going into the markets correspondingly falls
as actors approach their optimum portfolios. It is perhaps interesting that
these two effects more-or-less exactly cancel, leading to a stable price
throughout the trading period. The detailed mechanism ensuring this is
not obvious, but it can be argued that over a protracted trading period
no other possibility would be consistent.

Having constructed a model with stable prices, we can study the distribution 
of price fluctuations. Table \ref{fluctuation_params} shows the effects
of different model parameters on the size of the price fluctuations about the
mean, when an ensemble of simulations is sampled between time 2.25 and 2.5
trading periods.
Enough simulations are run to sample $10^7$ price pairs.

Figure \ref{log_pdf} shows the probability density function for 
price fluctuations about the mean, together with a lognormal distribution
with the same log-mean and log-standard-deviation, for comparison.
We see evidence for algebraic (otherwise known as `fat') tails in the 
time-series for price. This phenomenon is widely seen in financial
markets, and has been modeled in the past using the non-linear Fokker-Plank
equation \cite{Michael}.

\begin{figure}
\begin{center}
\includegraphics[width=0.85 \columnwidth]{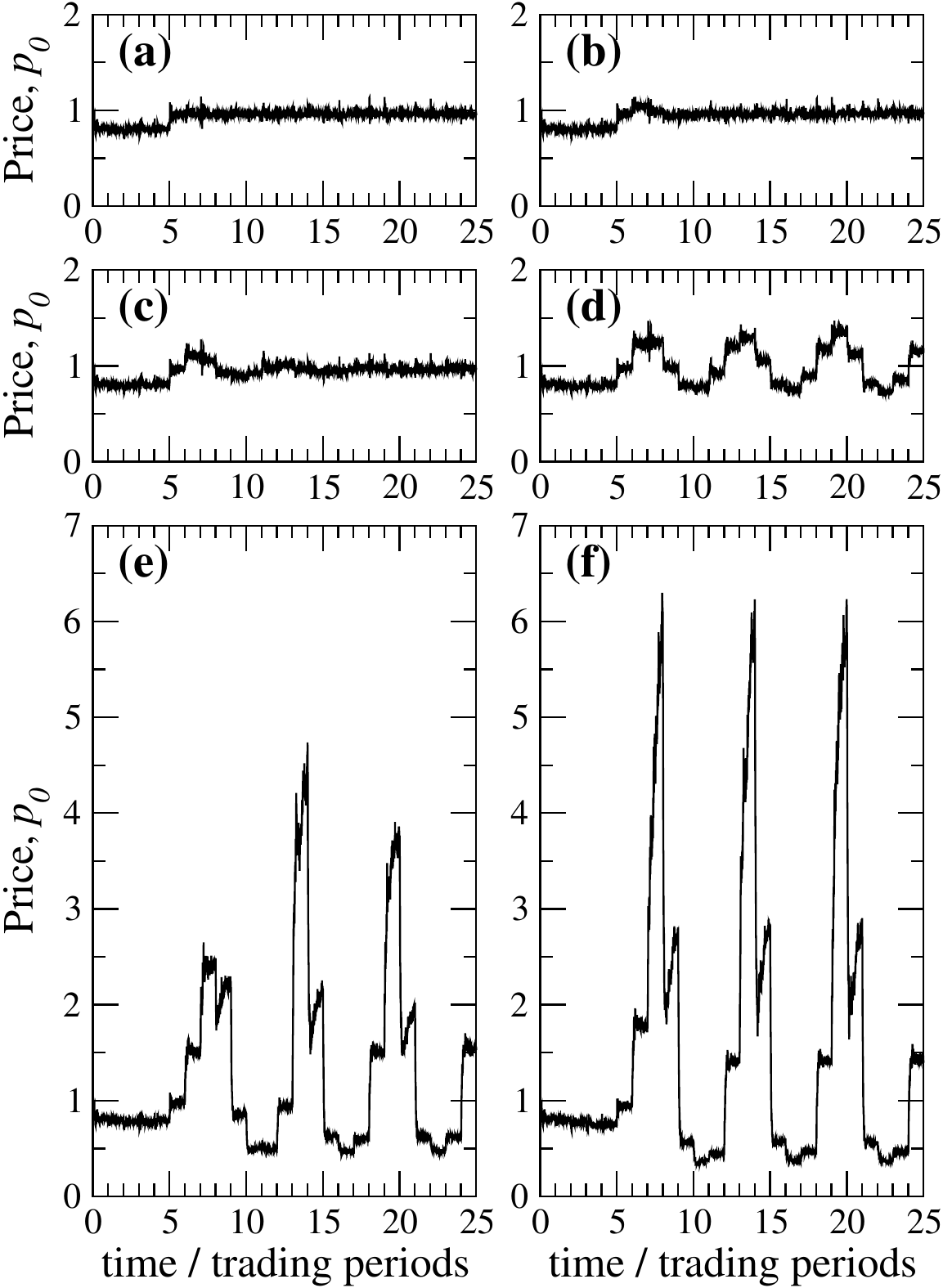}
\end{center}
\caption{Price (averaged over 10 turns) of good 0 versus time 
(in trading periods; each trading period 
being 4 rotations), for a simulation of initially 1000 actors,
half of whom produce unit amount of good $0$ before each trading period, and 
half good $1$ All goods are equally desirable to all actors, $d_{a,g}=1$. 
There are also 200 `inactive' actors, who have no money
and do not produce.
All have time preference parameter $f=3/2$. At a time of
5 trading periods, the inactive actors are given money but still have no 
ability to produce.
(a) None of the actors take into account inflation. (b)-(f) respectively
10\%, 20\%, 30\%, 40\% and 50\% of the actors take into account inflation. 
\label{inflation}}
\end{figure}

\section{Inflation expectations}
Many things can happen in an economy to affect the prices of goods.
For example, production of one or more goods may be restricted, new
technology may be introduced which renders some product obsolete at its
current price, or taxes may be imposed. These changes could strongly
affect the price of certain goods relative to others. However, a common
way for governments to appropriate saved labor is to increase the money
supply. This can be done in various ways. The most famous classical story was
of Dionysius the Elder of Syracuse\cite{Bullock}
who, having accumulated enormous debts, forced his subjects to bring their
one-drachma coins to him, which he then re-stamped as two drachmae.
Most subsequent (effective) increases in the money supply, such
as fractional reserve lending or the creation of a new tier of the money supply,
have been less explicit,
and are generally argued at the time to be temporary (and individually,
they might be). 

The ultimate consequence of an increase in the money supply is a rise in the
general price level; although with complex webs of industrial
input and output, the price rises of individual goods will have leads and
lags, generally described as the `Cantillon effect'\cite{Cantillon}.
When the prices of goods are generally perceived to be rising (termed
`price inflation') people need to take this into account when planning for
the future. Such `inflation expectations' are believed to have profound
effects on the dynamics of the economy.

In our next model (`Model 3') we include an awareness of inflation in the
psychological model of the actors, and then introduce an increase of 
the money supply in a very simple manner, by suddenly introducing new
actors into the simulation who arrive with no goods and no ability to
produce, but initially with some money. A record is kept of the average
price of each good over the previous trading period, which we write as
$\langle p_g(t-1)\rangle$, and also the average over the preceding 
trading period, which we write as $\langle p_g(t-2)\rangle$. 
Although price changes may be different for
different goods, we will convert these into a general price level,
taken here to be the geometric mean of the average price for
different goods.  The actors are then aware of the general
amount of price inflation, which we define as
\begin{equation}
I_p = \left[
\left. \prod_g \langle p_g(t-1)\rangle \right/ \prod_g \langle p_g(t-2)\rangle
\right]^{1/G}.
\label{I_p}
\end{equation}
The only effect relative to model 2, is that in Eqs.\ (\ref{Omega2phi})
and (\ref{phiopt}), the time preference parameter must be updated
to discount the expected future price inflation:
\begin{equation}
f_a \mapsto f_a / I_p .
\label{modified_f}
\end{equation}
The key difference is that whereas $f_a$ is fixed, $f_a/I_p$ has its
own dynamics.

In Model 3, if {\em all} actors are sensitive to inflation
and adapt their behavior to it using Eq.\ (\ref{modified_f}), then the
system has very large non-sinusoidal price oscillations.
However, if only a fraction of the actors are influenced by inflation
expectations, and there is a perturbation in the system (modeled in 
Figure \ref{inflation} as a sudden injection of new money), then the result is 
more moderate price oscillations. 
As the fraction of sensitive actors increases, the
oscillations increase in period. If no actors are sensitive to price inflation,
there is a step up in prices, complete within a single rotation, and then
stability returns. For small fractions of price-inflation-sensitive
actors, we find the oscillations are damped and stability is eventually 
reached;
while for larger fractions the larger resulting oscillations 
do not die out, and acquire more complex dynamics, with higher-frequency 
components. For the parameters used to generate Figure \ref{inflation}, 
the critical fraction for a change from damped to persistent oscillations 
is around 30\% of the actors.

The reason for the accelerating rise in price is that once the price rises have
started, those sensitive to inflation expect the future price to be even higher,
so dedicate more of their money to purchasing for immediate consumption.
They may potentially completely ignore the future (if $f_a/I_p < 1$),
and we are left with a situation similar to the hyperinflation of
Model 1. The reason for the return to lower prices has not been investigated
yet, but we speculate that it is due to the balance between the rate of growth
of expected prices through Eq.\ (\ref{I_p}) and a slower dynamic of the
actual (hyperinflationary) price increases.

\begin{figure}
\begin{center}
\includegraphics[width=0.85 \columnwidth]{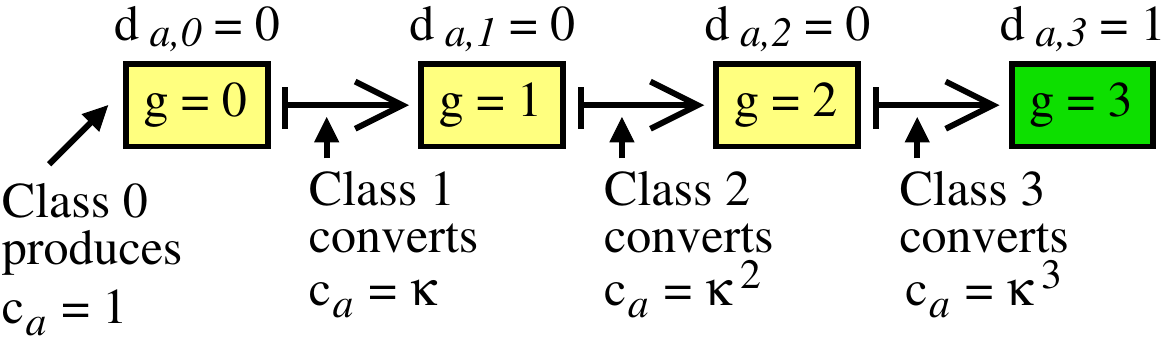}
\end{center}
\caption{An input-output model for a chain of 4 goods ($N_{\rm ch}=4$). 
There are four classes
of actors. Actors in class 0 produce unit amount of good $0$ at the start of
each cycle. Actors in class $1$ are able to convert up to $\kappa$ units of 
good $0$ into the same amount of good $1$ during a production period.
Similarly, Class 2 actors can convert up to $\kappa^2$ units of good $1$ 
into good $2$ and class 3 actors con convert up to $\kappa^3$ units of 
good $2$ into good $3$. The `competition factor' $\kappa\ge 1$ drives 
competition for goods earlier in the chain.
Only good $3$ has consumption value ($d_{a,3}=1$), while the others have
desirability $d=0$.
\label{io}}
\end{figure}

\begin{figure}
\begin{center}
\includegraphics[width=0.85 \columnwidth]{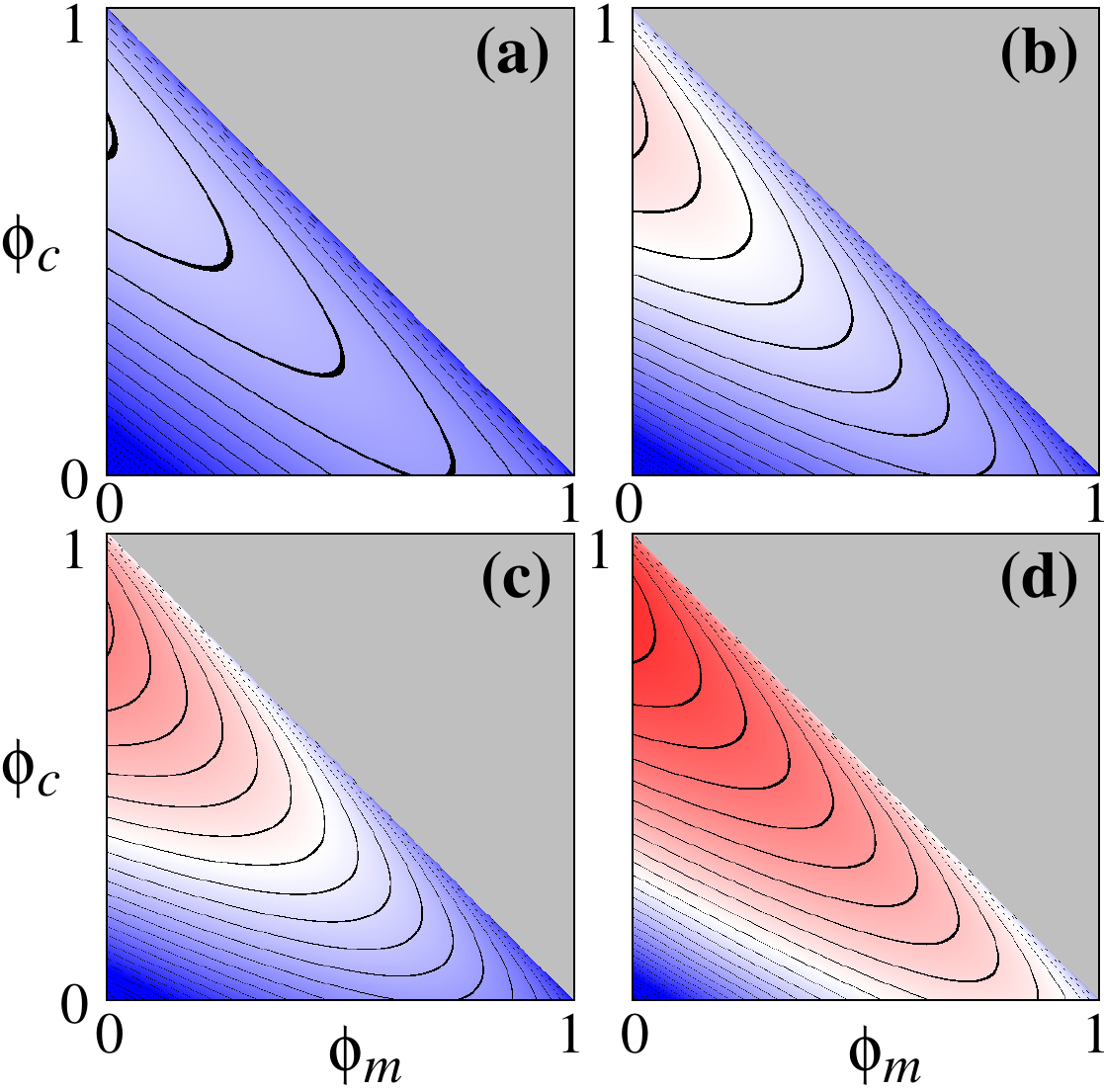}
\end{center}
\caption{Contour plots of the function $F = (1-\phi_m-\phi_c)^{1/2}
+f (\phi_m +\phi_c p_\delta/p_\gamma)^{1/2}$ which is proportional
to $\Omega_a^{[M4]}$ in Eq.\ (\ref{Omega4phi}).
(a) $f=1.5$, $p_\delta/p_\gamma = 1.2$. 
(b) $f=1.5$, $p_\delta/p_\gamma = 1.5$. 
(c) $f=1.5$, $p_\delta/p_\gamma = 1.8$. 
(d) $f=1.8$, $p_\delta/p_\gamma = 1.5$. 
Contours are every 0.025 units. White: $F=2$. Blue: $F<2$. Red: $F>2$
\label{contour_Omega}}
\end{figure}

\begin{figure}
\begin{center}
\includegraphics[width=0.95 \columnwidth]{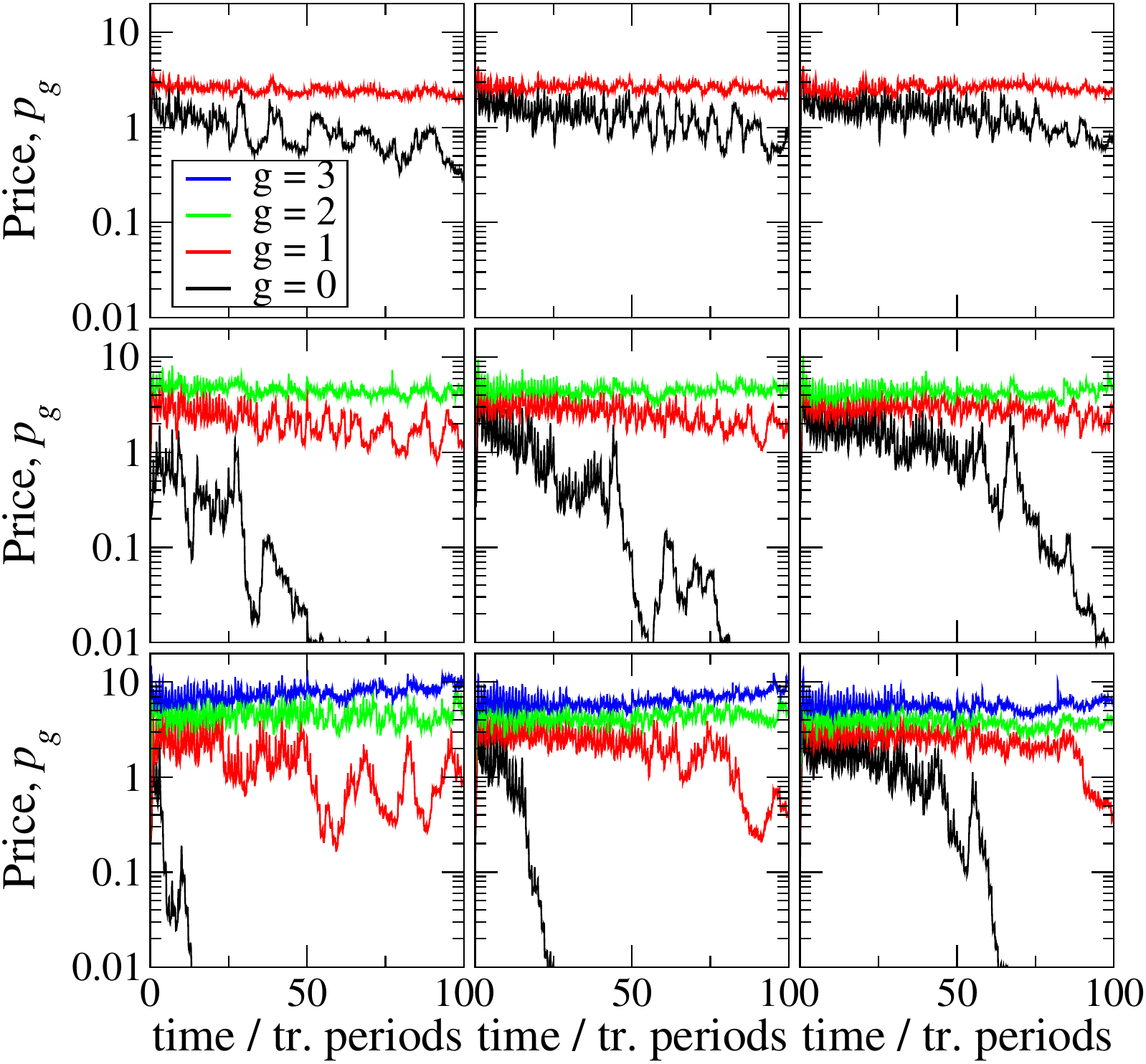}
\end{center}
\caption{Model 4 simulated over 100 trading periods, with 200 actors
in each class. Top row: a chain
of $N_{\rm ch}=2$ goods. Middle row: $N_{\rm ch}=3$. Bottom row: $N_{\rm ch}=4$.
Left column: competition factor $\kappa=1.2$. Middle column: $\kappa=1.5$.
Right column: $\kappa=2$. 
Figure \ref{io} shows the case $N_{\rm ch}=4$. 
Each point is an average price over 100 turns.
\label{io_sim1}}
\end{figure}

\begin{figure}
\begin{center}
\includegraphics[width=0.85 \columnwidth]{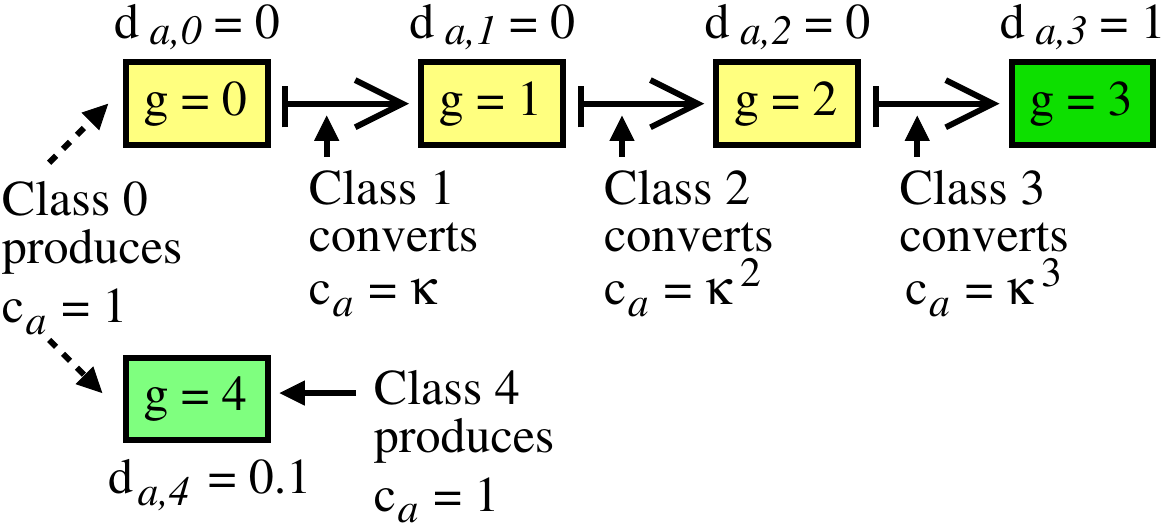}
\end{center}
\caption{Model 5: The input-output model of Figure \ref{io} modified to include
a new `subsistence good' 4, which now has much less desirability
than the final output (good 3) of the production chain, which is of
length $N_{\rm ch}=4$. There is an extra class of actors who produce
good 4, but actors in class 0 can switch to production of good 4 from
good 0 (or back) at the end of a turn, and will wish to do the former 
if $p_4 > p_0$. The probability of such an actor
switching, if it is advantageous to do so, is taken to be 10\%.
\label{io2}}
\end{figure}

\begin{figure}
\begin{center}
\includegraphics[width=0.85 \columnwidth]{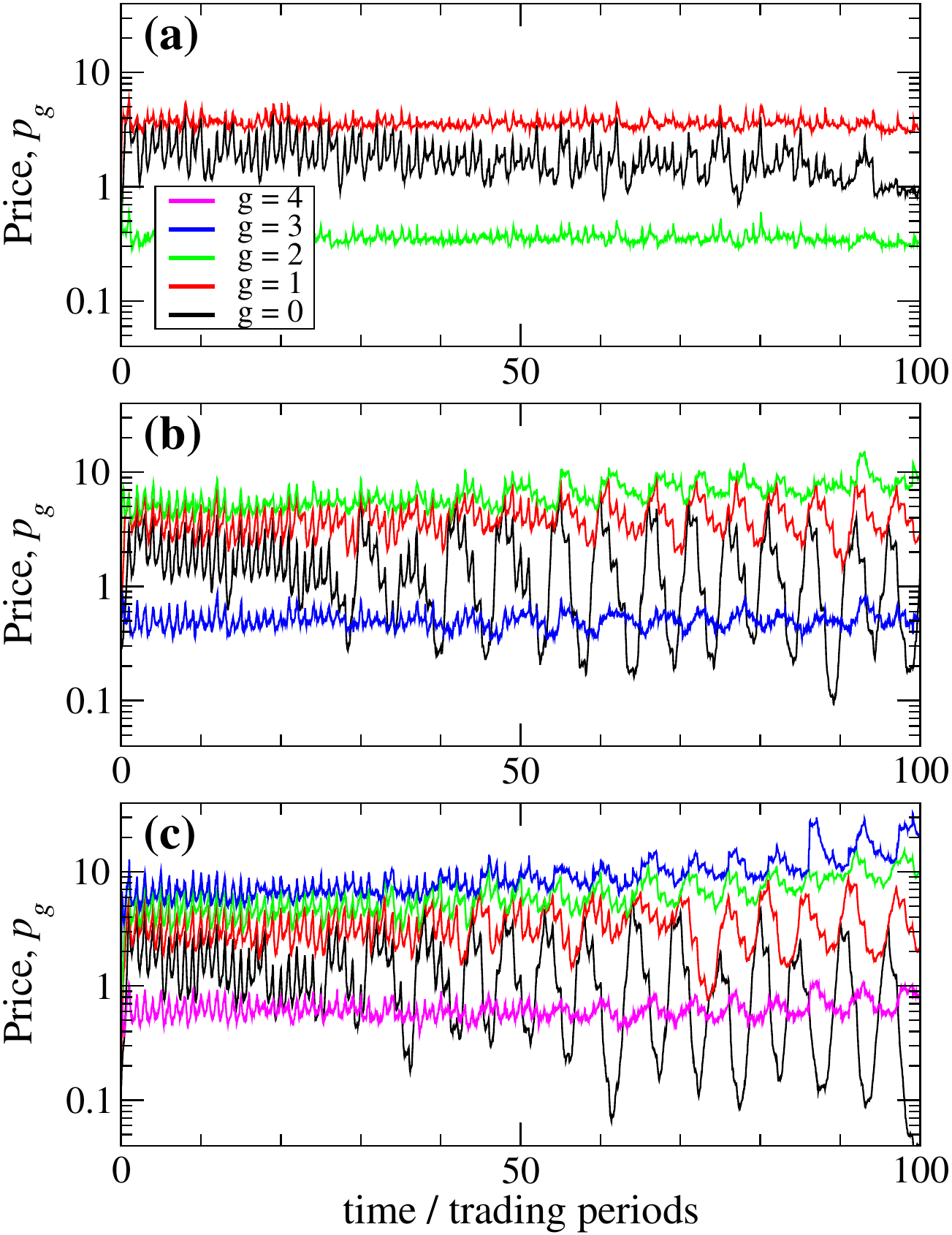}
\end{center}
\caption{Model 5 simulated over 100 trading periods, with 200 actors
in each class. (a) A chain
of $N_{\rm ch}=2$ goods. (b) $N_{\rm ch}=3$. (c) $N_{\rm ch}=4$.
In all cases the competition factor $\kappa=1.2$. 
Figure \ref{io2} shows the case $N_{\rm ch}=4$. 
Each point is an average price over 100 turns.
The probability of an actor in class 0 switching on a turn is 10\%, 
if advantageous to do so.
\label{switch_sim1}}
\end{figure}

\begin{figure}
\begin{center}
\includegraphics[width=0.85 \columnwidth]{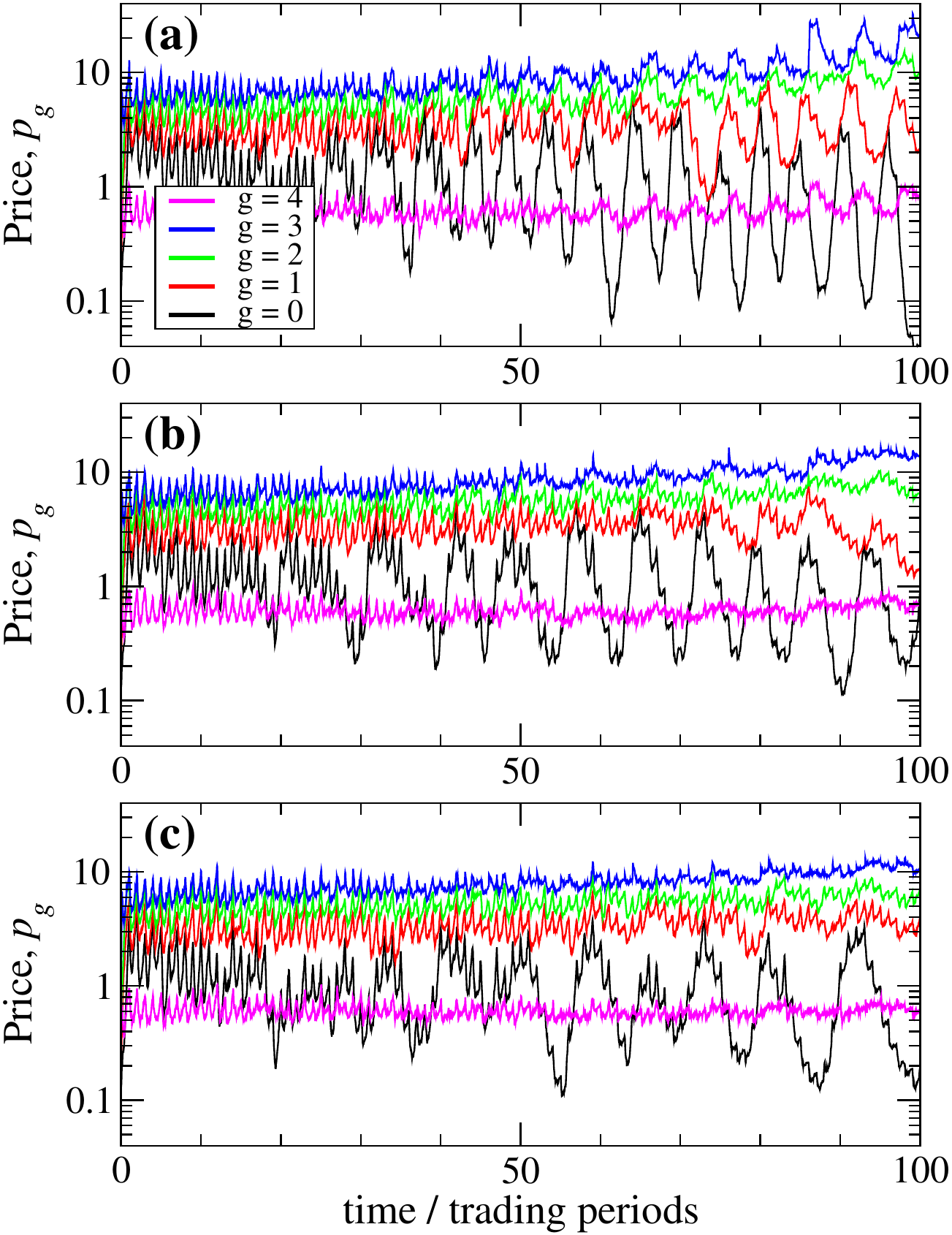}
\end{center}
\caption{Model 5 simulated over 100 trading periods, with 200 actors
in each class and a production chain of $N_{\rm ch}=4$. 
(a) The same simulation as Figure \ref{switch_sim1}(c),
where each turn actors in class 0 have a 10\% chance of being allowed
to switch production from good $0$ to good $4$.
(b) The chance is reduced to 3\% per turn. (c) The chance is reduced again,
to 1\%.
\label{switch_sim2}}
\end{figure}

\section{Input-output models}
Starting with the work of Leontief \cite{Leontief}, economies have been
modeled by constructing the network of inputs to different industries,
and outputs from them (which in turn may be inputs to further industries).
The framework we are presenting here can be extended to model such networks
(although conceptually it is currently only suited to `cottage industries'
where individuals manufacture, with relatively negligible capital requirements).

To do this, we posit the existence of a new class of actors. Previously
we have considered what we term `producers' who, at the start of a cycle,
are able to produce an amount of some good $\gamma$ {\it de novo}. 
We now add the
concept of an actor who is a `converter', and during the production
period is able to convert some maximum amount $c_a$ of a good in her possession
into another good. Typically, we would expect intermediate goods to have
no consumption value (so we set $d_{a,g}=0$ for these goods); 
but they should acquire a price through the same
mechanism that money acquires a value, namely that there is demand from
some subset of the actors. Figure \ref{io} shows schematically an
input-output model in our framework, with one class of producer actors and
three classes of `converter' actors.

To define our third model, we need to construct the utility function
which a `converter' actor tries to optimize. Suppose that such an actor
$a$ is able to convert up to a volume $c_a$ of good $\gamma$ into the same
volume of good $\delta$, during the production period that follows the
current trading period. We assume that no goods from the previous trading 
period can be stored into the
next trading period. The question $a$ faces is, `how much of good $\gamma$
to accumulate during the current trading period in her optimum portfolio,
versus retaining money?', since these are the only two means available to 
carry value through into the next trading period.

To answer this question, $a$ needs to choose how to split the current
liquidation value of her portfolio into three portions: a fraction
$\phi_m$ will be retained as money; a fraction $\phi_c$ will be used to buy 
good $\gamma$ for the purposes of conversion; and this leaves a fraction
$(1-\phi_m-\phi_c)$ which will be traded to obtain the optimum mix of
goods for consumption during the upcoming consumption period.
The utility function for `Model 4' (shown as a contour plot for some example 
cases in Figure \ref{contour_Omega}) is therefore:
\begin{eqnarray}
\Omega_a^{[M4]} &=& 2
\left\{
\left[ (1-\phi_m-\phi_c) L_a\right]^{1/2} +
\phantom{\left[ \left(\frac{x_x}{y_y} \right)\right]^{1/2}} 
\right.
\nonumber \\
&\ & \left.
\frac{f_a}{I_p} \left[ \left(\phi_m +\phi_c \frac{p_\delta}{p_\gamma}
\right)L_a \right]^{1/2}
\right\} 
\sqrt{ \sum_{g} \left( \frac{ d_{a,g}^2 }{p_g} \right) } ,\ \ 
\label{Omega4phi}
\end{eqnarray}
subject to the constraint that
\begin{equation}
\phi_c\, L_a/p_\gamma < c_a.
\label{phic_limit}
\end{equation}
If $p_\delta \le p_\gamma$ then it is not worthwhile for $a$ to convert
anthing, and using Eq.\ (\ref{phiopt}) we see that the solution is,
\begin{eqnarray}
\phi_c &=& 0 \\
\phi_m &=& \frac{(f_a/I_p)^2}{1+(f_a/I_p)^2}.
\end{eqnarray}
On the other hand, if $p_\delta > p_\gamma$ then $a$ would like to convert 
as much as 
possible, up to the limit set by Eq.\ (\ref{phic_limit}). If that limit is
not reached, $\phi_m = 0$ and we can again write down the solution by analogy
to Eq.\ (\ref{phiopt}). We then need to cover the case of the limit being
reached, so we define three new variables
\begin{eqnarray}
\phi_{\rm max} &\equiv& c_a \, p_\gamma / L_a  
\\
\phi_a &\equiv& 
\frac{(f_a/I_p)^2(p_\delta/p_\gamma)}{1+(f_a/I_p)^2(p_\delta/p_\gamma)}
\\
\phi_b &\equiv& \frac{(f_a/I_p)^2 (1-\phi_{\rm max}) 
- (p_\delta/p_\gamma)\phi_{\rm max}}{ 1+(f_a/I_p)^2} 
\end{eqnarray}
and the solution is
\begin{eqnarray}
\phi_c &=& \min \left( \phi_{\rm max} , \phi_a \right)
\\
\phi_m &=& \left\{
\begin{array}{ll}
0 & \phi_a < \phi_{\rm max} \\
\max \left( \phi_b , 0 \right) & \phi_a \ge \phi_{\rm max}
\end{array}
\right. .
\end{eqnarray}
This sets the optimum portfolio. After each turn, we note the amount of the 
input good $\gamma$ that $a$ is intending to save for future conversion.
At the end of a trading period, $a$ may not have successfully bought all
of this, so the amount converted is 
the minimum of what $a$ intends from the calculation above, and
the amount of good $\gamma$ that $a$ actually possesses.

Figure \ref{io_sim1} shows the results for different chain lengths,
$N_{\rm ch}$. To encourage price formation early in the production chain, 
we have also introduced a `competition factor' $\kappa$, which is the ratio of 
conversion capacity for an actor, compared to an actor earlier in the 
production chain (see Figure \ref{io}). Even with this expedient, we see
that stable price formation does not occur, with the first failure
mode being the price of good $0$, at the base of the production
chain, falling over time (with some delay, if $\kappa$ is large).
Prices further along the production chain then start to fall.

We hypothesize that the underlying reason for this is price inelasticity
for good $0$. No matter how low the price, actors in class 0 will
continue to produce, at their maximum capacity $c_a = 1$.
To test this hypothesis, we modify the model to `Model 5'.
In this new version, there is a `subsistence good' (good 4 in the case
$N_{\rm ch}=4$), which is
produced by a new class of actor, and has a relatively small desirability
($d_{a,4}=0.1$)
for everyone compared to the product at the end of the chain.
Actors in class 0 have
the ability to switch production between good $0$ and the subsistence
good $4$, depending on their relative price (see Figure \ref{io2}). 
The hypothesis is that this will provide a floor for the price of good $0$, 
and therefore aid in stable price formation. Specifically, every turn
we allow an actor in class 0 a 10\% chance of switching production
type. Since there are 4 rotations in a trading period, such an
actor will typically have 4 turns before a production period.

Figure \ref{switch_sim1} shows that Model 5 does limit how small the 
price of good $0$ can become, but leads to a new mechanism for large price
oscillations. Those oscillations appear to become unstable, at least for 
$N_{\rm ch} > 2$, which means that the present model is unable to
produce price stability when supply chains are non-trivial. 
Figure \ref{switch_sim2} shows he effect of reducing the probability of a
switch in production by actors in class 0. Slightly greater stability
is seen (in terms of a slower inflation in the price of good $4$), and
a longer period of price oscillations results.

\section{Discussion}
In this paper, we have constructed simple utility functions for market actors.
Because we are modeling the psychology of individual actors, which
is the foundation of all real economic activity, we refer
to this approach as {\it `ab initio'}.

These utility functions can be extended in many ways, but once specified,
rational strategies to optimize these functions follow naturally
(although with some ambiguity). This makes models based on utility functions
applicable to a wide range of scenarios and (when combined with market
rules) produce interesting emergent behavior. Here we have shown
`fat tails' in price fluctuations, and price oscillations as examples
of such emergent behavior.

Algebraic tails in price sequences have long been remarked upon, and are
broadly understood in terms of statistical processes.
However, we believe this work is the first time they have been shown to
emerge from a plausible model of interacting rational agents.

The existence of cycles in the economy has been known since the work of 
Sismondi \cite{Sismondi} in the eighteenth century and later Samuel 
Benner \cite{Benner} 
in the nineteenth. Many theories have subsequently been created
to account for oscillations of prices and production in the economy, 
which involve mechanisms not present in our simple `Model 3', so it is 
interesting that oscillatory behavior can arise with only minimal assumptions
-- in this case just a rational response of actors to forward projection of 
past price inflation. In practice, more complex processes could well be 
responsible for the observed economic cycles in the real world.

Economies -- even the `cottage economies' of individual manufacturers --
involve production of several goods, some of which are feed-stocks for others. 
When we attempt to model these networks of interactions (usually
captured in economic theory as `input-output' models), we find that
stable price formation is difficult to achieve. In other words, any long
supply chains would have to be `in-housed' in this model. However, 
there is scope for including more of the complexities of real
economies within the broad framework we have presented here.
Some of those added complexities may allow more realistic behavior
to emerge, and we hope cross-disciplinary efforts will bring the
necessary insights.

\acknowledgements

Source code, in the C programming language, used to run the simulations
and generate the data for the figures in this paper are hosted at
`Source Forge':
\href{https://sourceforge.net/projects/ab-initio-mkt/files/}{https://sourceforge.net/projects/ab-initio-mkt/files/}

\end{document}